\documentclass[useAMS,usenatbib]{mn2e} \input{epsf}

\newif\ifAMStwofonts %\AMStwofontstrue

\def\lesssim{\mathrel{\hbox{\rlap{\hbox{\lower4pt\hbox{$\sim$}}}\hbox{$<$}}}}
\def\gtrsim{\mathrel{\hbox{\rlap{\hbox{\lower4pt\hbox{$\sim$}}}\hbox{$>$}}}}
\def\msun{${\rm M}_{\odot}$~}
\def\msol{${\rm M}_{\odot}$}

\def\l_lsun{$\log{L/\rm L_{\odot}}$~}
\def\masa_msun{$M/ \rm M_{\odot}$~}

\def\m_mstar{$M/M_{*}$~}

\def\aap{A\&A}
\def\aj{AJ}
\def\apj{ApJ}
\def\apjl{ApJ}
\def\apjs{ApJS}
\def\mnras{MNRAS}

\def\pasp{PASP}
\def\nat{Nature}

\title[White dwarf evolution in PSR~J1713+0747 system]{The evolutionary status
of the white dwarf companion of the binary pulsar PSR~J1713+0747}

\author[O.~G. Benvenuto, R.~D. Rohrmann \& M.~A.   De   Vito]
{
O.~G. Benvenuto$^{1,2}$\thanks{Member of  the Carrera del Investigador
Cient\'{\i}fico, Comisi\'on de  Investigaciones Cient\'{\i}ficas de la
Provincia    de    Buenos    Aires    (CIC),    Argentina.     Emails:
obenvenu@astro.puc.cl; obenvenuto@fcaglp.unlp.edu.ar}
R.~D.  Rohrmann$^{3}$\thanks{Member of  the  Carrera del  Investigador
Cient\'{\i}fico, Consejo  Nacional de Investigaciones Cient\'{\i}ficas
y T\'ecnicas (CONICET), Argentina.  Email: rohr@oac.uncor.edu}
M.~A.  De Vito$^{2}$\thanks{Fellow of the CIC; Universidad Nacional de
La Plata (UNLP); Instituto  de Astrof\'{\i}sica de La Plata (CONICET).
Email: adevito@fcaglp.unlp.edu.ar} 
\\
$^{1}$ Departamento de Astronom\'{\i}a y Astrof\'{\i}sica, Pontificia
Universidad Cat\'olica, Vicu\~na Mackenna 4860, Casilla 306, Santiago, Chile\\
$^{2}$ Facultad de Ciencias Astron\'omicas y Geof\'{\i}sicas, Universidad
Nacional de La Plata, Paseo del Bosque S/N, B1900FWA, La Plata,\\ Argentina\\
$^{3}$ Observatorio Astron\'omico, Universidad Nacional de C\'ordoba, Laprida
854, 5000 C\'ordoba, Argentina }

\begin{document}

\date{March 17}

\pagerange{\pageref{firstpage}--\pageref{lastpage}} \pubyear{2005}

\maketitle \label{firstpage}

\begin{abstract}

Recently Splaver and  coworkers have measured the masses  of the white
dwarf  and the neutron  star components  of the  PSR~J1713+0747 binary
system  pair by  means of  the  general relativistic  effect known  as
Shapiro Delay with very high  accuracy. Employing this data we attempt
to  find  the original  configuration  that  evolved  to the  observed
system.   For  this purpose  we  perform  a  set of  binary  evolution
calculations trying  to simultaneously account for the  masses of both
stars  and the  orbital  period.  In doing  so,  we considered  normal
(donor) stars with an initial mass of 1.5~\msol, while for the neutron
star  companion  we  assumed  a  mass of  1.4~\msol.  We  assumed  two
metallicity values for the donor  star ($Z = 0.010$ and $0.020$)
and that the initial orbital period was near 3 days. In order to
get a good agreement between the masses of the models and observations
we had  to assume that  the neutron star  is only able  to retain
$\lesssim~0.10$   of   the    matter   transferred   by   the   donor
star.  Calculations were  performed  employing the  binary hydro  code
developed by Benvenuto \& De  Vito (2004), that handles the mass
transfer rate  in a fully implicit way  together with state-of-the-art
physical  ingredients  and diffusion  processes.   Now  our code  also
includes a  detailed non-grey treatment  for the atmospheres  of white
dwarfs.

We  compare the  structure  of  the resulting  white  dwarfs with  the
characteristic  age of  PSR~J1713+0747 finding  a nice  agreement with
observations by  Lundgren et  al. especially for  the case of  a donor
star with  $Z= 0.010$.  This result indicates  that, at least  for the
purposes  of this paper,  the evolution  of this  kind of  binary 
system is fairly well understood.

The models predict that, due to diffusion, the atmosphere of the white
dwarf is an almost hydrogen-pure one. We find that such structures are
unable to  account for  the colours measured  by Lundgren et  al. 
within their error bars. Thus, in spite of the very good agreement of
the model  with the main characteristics  of the system,  we find 
that some discrepances in the white dwarf emergent radiation remain to
be explained.

\end{abstract}


%---------------------------------------------------------------------
\section{Introduction} \label{sec:intro}

PSR~J1713+0747  was discovered  in  a survey  for millisecond  pulsars
(MSPs) with  the 305~m Areceibo radio telescope  (Foster, Wolszczan \&
Camilo 1993). It  has been found that the pulsar has  a spin period of
4.57  ms.   Systematic changes  in  the  apparent  pulsar period  have
indicated the binary nature of  this object. Foster et al. (1993) made
the   first  determination   of   the  binary   model  parameters   of
PSR~J1713+0747. They  found a  nearly circular, 67.8~day  binary orbit
with  a  low mass  white  dwarf  (WD)  companion. They  estimated  the
companion mass via the mass function. Assuming the canonical value for
the mass of the NS ($M_{NS}=1.4$~\msol) they set a lower limit of for
the   companion   mass   of   $M_{WD}  \geq   0.28$~\msol.    Clearly,
PSR~J1713+0747 belongs  to the family of neutron  stars (NSs) recycled
by  mass  and  angular  momentum  transfer  from  its  normal  (donor)
companion (Alpar et al. 1982).  This process leads to the formation of
a MSP together  with a cool WD companion. For a  recent review on this
kind of binary systems see Stairs (2004).

PSR~J1713+0747  has been  continuously  observed  with an
increasing degree of  accuracy, which has been large  enough to reveal
the general  relativistic effect known  as Shapiro delay.  This occurs
due to a  slight change in the arrival time of pulsar radiation
due  to the  curvature of  space near  the WD  companion of  the MSP.
According to general relativity, pulses are retarded as they propagate
through  the  gravitational  potential  well  of the  WD.   The  first
detection of Shapiro delay in PSR~J1713+0747 is due to Camilo, Foster,
\& Wolszczan (1994). They were able  to set lower limits of the masses
of the WD ($M_{WD}>$~0.27~\msol) and NS ($M_{NS}>$~1.2~\msol).

In a very recent work, Splaver et al. (2005) have reported the results
of  twelve years of  observations of  PSR~J1713+0747. The  timing data
yielded a large  improvement in the measurements of  the Shapiro delay
that allowed for  the determination of the individual  masses for both
components of the system.   In principle Shapiro delay yields $M_{WD}$
and $\sin{i}$, where $i$ is the inclination angle. In practice, unless
$\sin{i}  \sim  1$, Shapiro  delay  is  highly  covariant with  $a_{1}
\sin{i}$ \footnote{$a_{1}$  is the  orbital semi-major axis.}   in the
timing  model fit and  hence difficult  to measure.   Nevertheless, as
shown by Splaver  et al (2005), in the  case of PSR~J1713+0747 Shapiro
delay  can be clearly  distinguished. These  high-precision detections
allow  for the  masses of  the  pulsar and  the companion  star to  be
separately measured,  being $M_{NS}=  1.3 \pm 0.2$~\msun  and $M_{WD}=
0.28  \pm  0.03$~\msol.   These  authors  repeat the  statistical
analysis  of  timing  solutions,  now combined  with  the  theoretical
orbital period-core mass  relation, and they obtain a  larger mass for
the NS: $M_{NS}= 1.53^{+0.08}_{-0.06}$~\msun (68 \% confidence).

The pulsar masses in  systems that evolve into pulsar-WD binaries
are expected to  be larger that those in NS-NS  binaries (in this last
case,   the   pulsar   masses    are   expected   within   the   range
1.25-1.44~\msol).   This  is  because  the systems  that  evolve  into
pulsar-WD  systems undergo extended  periods of  mass transfer.   In a
binary population  study performed by Thorsett  \& Chakrabarty (1999),
the  authors  find a  narrow  Gaussian  distribution  with a  mean  of
1.35~\msun  and a  width  of  0.04~\msol. The  mass  derived when  the
orbital period-core mass relation is imposed in PSR J1713+0747 implies
a significantly heavier NS.  Apart from PSR~J1713+0747 there are other
MSP~+~WD systems for which it has been possible to measure the Shapiro
delay and  the masses of  each component of  the pair. One of  them is
PSR~J0437-4715 (van Straten et al.  2001) for which $M_{WD}= 0.236 \pm
0.017$~\msol,   and   $M_{NS}=   1.58\pm0.18$~\msol.  The   other   is
PSR~B1855+09  for   which  $M_{WD}=0.258^{+0.028}_{-0.016}$~\msun  and
$M_{NS}=  1.50^{+0.26}_{-0.14}$~\msun for  the pulsar  mass  (Kaspi et
al. 1994).  The uncertainties on  all these measurements are large and
the question  of the distribution  of pulsars masses in  these systems
remains open.

One very important parameter is the characteristic age of the pulsar 
$\tau_{PSR}$. The age of a spin-down powered pulsar is given by

\begin{equation}
\tau_{PSR} = \frac{P}{(n - 1)\dot{P}}
          \biggl[1 - \biggl(\frac{P_{0}}{P}\biggr)^{n-1}\biggr]
\end{equation}

\noindent  where $P$  is  the  pulsar period,  $\dot  P$ the  temporal
derivative of  the period, $P_{0}$ is  the initial period  and $n$ the
braking index. For magnetic dipole  radiation we have $n=3$, and if we
assume $P_{0} \ll P$, the pulsar spin-down age is given by $\tau_{PSR}
= P  / (2  \dot P)$ ,  the ``characteristic  age'' of the  pulsar. The
uncertainty in the  pulsar age is caused by $n$  and $P_{0}$. On
the  other hand,  the  transverse  velocity of  the  pulsar causes  an
apparent  acceleration   that  contributes  to   the  observed  period
derivative.   Ignoring  this  latter  effect  can  lead  to  erroneous
estimations of  the characteristic pulsar age.  Splaver  et al. (2005)
found   for   PSR~J1713+0747,  $\tau_{PSR}   =   8$~Gyr.   Using   the
restrictions for the mass of the  WD in this system, Hansen \& Phinney
(1998) obtain  cooling ages  in the  range  6.3-6.8 Gyr,  but if  they
consider the  dispersion measure  distance estimates, they  can obtain
cooling ages up to 13.2 Gyr.

If the pulsar has a close  companion WD, we expect the cooling age for
the WD to  approximately match the age of the pulsar.  For the kind of
systems we  are here interested in, $\tau_{PSR}$  should correspond to
the age  of the helium WD  (hereafter He~WD) counted since  the end of
the  main (initial)  Roche Lobe  OverFlow (RLOF)  episode. This  is so
because the initial  RLOF is the only one able  to transfer the amount
of matter needed to spin  up the NS (probably $\approx 0.1$~\msol). As
we shall show  below, models of normal stars  leading to the formation
of the  WD companion suffer  from at least one  hydrogen thermonuclear
flash  leading to  a supplementary  RLOF. However,  these flash-driven
RLOFs  are not  relevant in  spinning up  the NS  because of  the tiny
amount of transferred  matter (Benvenuto \& De Vito  2004). Because of
these reasons,  the WD in PSR~J1713+0747  should have spent  a time of
$\tau_{PSR} \approx 8$~Gyr since the end of the main RLOF episode.

The aim of  this work is to  test the present status of  the theory of
stellar evolution in binary systems by computing a set of evolutionary
tracks  in  order  to   reproduce  the  main  characteristics  of  the
PSR~J1713+0747 system, particularly the  masses of the components, the
orbital period and the characteristic timescale for the cooling of the
WD. In our opinion this is  an interesting problem because of the high
precision of the measurements performed  for the system we are dealing
with.

The  WD companion  of PSR~J1713+0747  has been  detected  by Lundgren,
Foster \& Camilo (1996)  with Hubble Space Telescope observations (see
Table~\ref{table:observa}). This fact provides us with the interesting
possibility of  confronting theory with  observations. In doing  so we
shall  employ   the  available  observational   data  of  the   WD  in
PSR~J1713+0747  as  a  test   of  the  accuracy  of  our  evolutionary
calculations and {\it  not} as the main data to  account for. We shall
do so because of two reasons. First, we judge that the accuracy of the
radioastronomical  measurements of  the system  is so  high  that they
should  be   considered  as   the  fundamental  quantities   for  this
problem. Second, this  procedure is interesting in order  to gauge the
ability of  the binary  stellar evolution  theory  combined with
atmosphere models  in predicting the  characteristics of a WD  and to
evaluate the possibility  of helping for a first  optical detection of
WDs as companions of MSPs.

\begin{centering}
\begin{table}
\caption{\label{table:observa}   Main   characteristics   of  the   WD
component  of  the  PSR~J1713+0747  system,  taken  from  Lundgren  et
al. (1996).  The numbers in  parentheses represent the  uncertainty in
the last digits quoted.}
\begin{tabular}{lc}
\hline
Quantity & Measured value\\
\hline
\hline
$m_{B}$   & $>$27.1 \\
$m_{V}$   & 26.0(2) \\
$m_{I}$   & 24.1(1) \\
$B-V$     & $>$1.1 \\
$V-I$     & 1.9(2) \\
$m-M$     & 10.2(5) \\
$E_{B-V}$ & 0.08(2) \\
\hline
\end{tabular}
\end{table}
\end{centering}

Regarding  the  viability of  this  procedure,  we  should quote  that
recently, Benvenuto  \& De Vito (2005)  have been able  to account for
the   masses,  orbital   period,   and  characteristics   evolutionary
timescales of the WDs in PSR~J0437-4715 and PSR~B1855+09.

Our particular  interest on  this type of  systems has  been focused
onto the  study of  low mass  He~WDs (see Benvenuto  \& De  Vito 2004,
2005).  Besides,   these systems  are  appropriate   for  testing  our
evolutionary models of cooling He~WDs.

The   reminder   of  the   paper   is   organized   as  follows.    In
Section~\ref{sec:numerical}  we  present the  numerical  code we  have
employed.  In  Section~\ref{sec:results} we describe  the evolutionary
results we  have found for the PSR~J1713+0747  binary system. Finally,
in   Section~\ref{sec:discu}  we   discuss  the   implicates   of  our
calculations and summarize the main conclusions of this work.

%---------------------------------------------------------------------
\section{The Numerical Code} \label{sec:numerical}

The binary  stellar evolution  calculations presented below  have been
performed employing the stellar code described in Benvenuto \& De Vito
(2004) now with  a non-grey model  atmosphere as detailed  in Rohrmann
(2001). We briefly mention that  our code employs a generalized Henyey
technique that allows for the computation of mass transfer episodes in
a fully  implicit way (Benvenuto  \& De Vito  2003).  The code  has an
updated description of opacities, equation of state, nuclear reactions
and  diffusion  while  we  simultaneously  compute  orbital  evolution
considering  the  main processes  of  angular  momentum loss:  angular
momentum   carried  away  by   the  matter   lost  from   the  system,
gravitational radiation, and magnetic braking. To be more specific, in
accounting for the magnetic braking we follow Podsiadlowski, Rappaport
\& Pfahl (2002) and  incorporate Eq.~36 of Rappaport, Verbunt \&
Joss (1983), assuming $\gamma=4$.  We neglect irradiation of the donor
star by the MSP.

In our treatment of the orbital evolution of the system, we shall
consider that  the NS  is able  to retain a  fraction $\beta$  of the
material coming  from the  donor star: $\dot{M}_{NS}=  -\beta \dot{M}$
(where $\dot{M}_{NS}$  is the accretion  rate of the NS  and $\dot{M}$
the  mass transfer  rate  from  the normal  star).  $\beta$ should  be
considered as a free parameter,  because at present it is not possible
to  compute  it  from  first  principles.\footnote{For  the  cases  of
PSR~J0437-4715  and   PSR~B1855+09,  in  order  to   account  for  the
characteristics of each of these systems, it has been found (Benvenuto
\&  De  Vito 2005)  that  $\beta$  should  be $\lesssim  0.12$.}   For
simplicity, it  will be considered   as a constant  throughout all
the RLOF  episodes. We shall assume  that the material  lost from the
binary  system  carries away  the  specific  angular  momentum of  the
compact object.

In  our previous  works on  binary evolution  we have  considered grey
atmospheres.  Now, for a proper  treatment of the cooling behaviour of
the  WD companion  of  the millisecond  pulsar,  we have  incorporated
detailed non-grey model atmospheres as an outer boundary condition for
evolutionary models  of the  WD. The use  of non-grey  atmospheres for
this purpose  is particularly  important in the  late stages  of He~WD
cooling (Serenelli et al. 2001). Besides, the atmosphere code allow us
to  determine the  emergent  flux from  the  WD and  to interpret  its
magnitudes and colours.

The procedures  employed in the  calculation of LTE  model atmospheres
are basically the same as described in Rohrmann (2001) and Rohrmann et
al.   (2002).    Models   are   computed  assuming   hydrostatic   and
radiative-convective  equilibrium.    Convective  transport  is
treated within  the mixing-length theory.  We  include line blanketing
by  Lyman  lines of  hydrogen  and  helium,  and the  pseudo-continuum
opacity  from   the  Balmer  and   Paschen  edges.   Collision-induced
absorption (CIA)  becomes an important  source of infrared  opacity in
WDs cooler than $T_{eff}\approx 5000$  K. Our code includes a detailed
description of CIA due to  H$_2$-H$_2$ (Borysow et al. 2001), H$_2$-He
(Jorgensen  et al.  2000)  and H-He  (Gustafsson  \& Frommhold  2001).
Chemical equilibrium is based  on the occupation probability formalism
as described in Rohrmann et  al. (2002), and includes H, H$_2$, H$^+$,
H$^-$, H$_2^+$,  H$_3^+$, He, He$^-$, He$^+$,  He$^{++}$, He$_2^+$ and
HeH$^+$.   Ionic  molecules  He$_2^+$  and HeH$^+$  were  incorporated
recently   (for details,  see  Gaur et  al.  1988  and Harris  et
al.  2004).  We  also  added  the continuous  absorption  opacity  of
He$_2^+$ following Stancil (1994).

In order to optimize the  numerical effort, initially we shall compute
the evolutionary sequences employing  grey atmospheres. In this way we
search for  the correct  configuration for the  binary system  able to
account  for the main  characteristics of  the binary  system (masses,
orbital period and characteristic  age). After identifying a plausible
configuration we have  at hand a complete evolutionary  track from the
Zero Age Main Sequence up to  the end of the WD cooling sequence. Then
we rerun the  code, now including the non-grey  stellar atmosphere. In
doing so we shall start this  detailed computation from the top of the
last  cooling track, near  the conditions  in which  the star  has its
absolute maximum effective temperature. In this way we avoid computing
a full  evolutionary track with a non-grey  atmosphere without missing
relevant  detail of  the  physics  of the  problem  we are  interested
in. This is very important in order to keep the numerical computations
affordable.

%---------------------------------------------------------------------
\section{Results} \label{sec:results}

\begin{figure*} \epsfysize=320pt 
\epsfbox{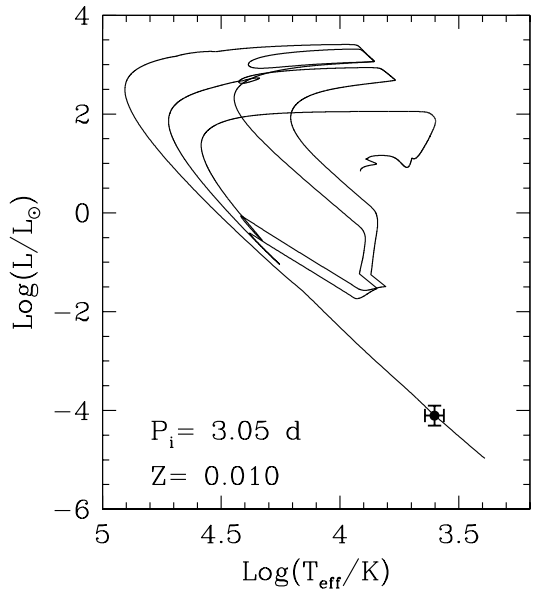}  
\epsfysize=320pt
\epsfbox{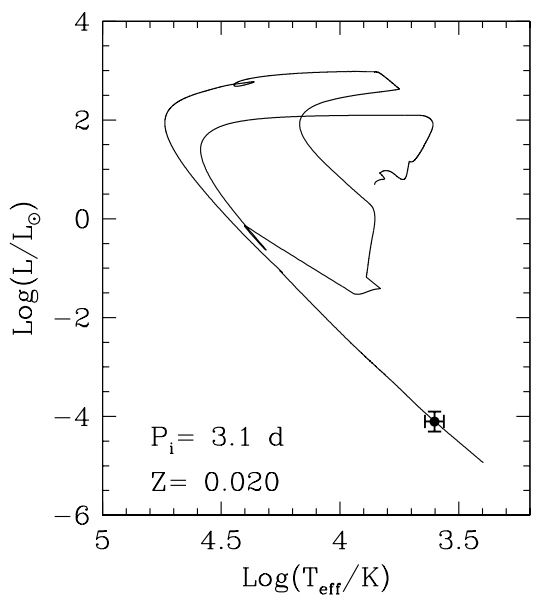} 
\caption{Evolutionary  tracks  for  stars  with  an  initial  mass  of
1.50~\msun  as companions  of  a NS  with  a canonical  mass value  of
1.4~\msol.  Left panel  corresponds to  the  case of  an abundance  of
$Z=0.010$ and an initial  period of 3.05~days. Right panel corresponds
to the  case of  solar chemical composition  and an initial  period of
3.10~days. For the  case of $Z=0.010$ the star  undergoes two hydrogen
thermonuclear flashes while for  $Z=0.020$ the object experiences only
one.  For comparison  we  also include  the  luminosity and  effective
temperature  for the  WD in  the  PSR~J1713+0747 as  deduced from  our
models with  a solid  circle with its  corresponding error  bars.  See
text for further details.} 
\label{Fig:tracks} 
\end{figure*}

In order to account for the main characteristics of the PSR~J1713+0747
binary system we have computed a set of evolutionary models. As stated
above, our  primary quantities are  the masses and orbital  period and
the observed WD conditions are employed  as a test of how plausible is
the  theory of  binary stellar  evolution we  are employing.  For this
purpose we  have to set values  for many initial  parameters. We shall
assume the  initial mass  of the  normal component of  the pair  to be
1.5~\msun  while for  the  NS  we employ  its  ``canonical value''  of
$M_{NS}=  1.4$~\msol.   Here we  shall  consider  two  values for  the
metallicity of the normal star: $Z= 0.010$ and $0.020$.

Regarding  mass  transfer, we  have  assumed  that $\dot{M}_2=  -\beta
\dot{M}_1$ where  subindex 1 (2)  corresponds to the  normal (neutron)
star.  As $\beta$ is  constant, this  implies that  $M_2 +  \beta M_1=
M^{i}_2  + \beta M^{i}_1$  where superscript {\it  i} indicates
that  these are  initial masses.  Immediately we  find $\beta=  (M_2 -
M^{i}_2)/(M^i_1  -  M_1)$. If  we  assume  that  the NS  has  accreted
$\lesssim 0.1$~\msol,  then $\beta \lesssim 0.10$.  Hereafter we shall
adopt the upper limit value of $\beta= 0.10$.

After  considerable  search  in  the  parameter  space  we  found  two
plausible solutions. These are a pair with the above mentioned masses,
a  metallicity of  $Z=0.010$ and  an initial  period  $P_i=3.05$~d and
another with  $Z=0.020$ and an initial period  $P_i=3.10$~d.  From
here  on we  shall explore  the evolution  of these  systems  and its
comparison with observations. While we have not performed (a very time
consuming) finer exploration in the parameter space, we interpret that
these two  systems are suitable  for our purposes.  In any case  it is
worth mentioning  that significantly lower  mass values for  the donor
star would require times in excess  of the Hubble time in order to get
a dim WD  object like the one observed for the  system we are studying
here. Also, longer initial periods would force the system to undergo a
common envelope  episode because at the  onset of the  RLOF, the donor
star would have a very  deep outer convective zone (hereafter OCZ) and
consequently to a considerable shrinkage of the orbit.

We decided not  to explore in detail the possibility  of a lower value
for the  metallicity. As it will  be clear below, the  systems we have
chosen  produce  WD  objects  with  mass values  compatible  with  the
($1\sigma$) error  bar but almost in  its upper limit. If  we look for
values of $Z  < 0.010$ we would find  a larger final WD mass  as it is
clear  from  the  mass-period  relation  given by  Nelson,  Dubeau  \&
MacCannell (2004)

\begin{equation}
\log{\bigg(\frac{P_f}{1 day}\bigg)}=
10.7 \bigg( \frac{M_f}{M_\odot} \bigg) + 0.3 \log{Z} - 0.98277.
\end{equation}

Thus, for lower $Z$-values from the ones considered here we would find
WD masses  in excess  from the observed  error bars.   This result
suggests  that  a  metallicity  lower  than  $Z  <  0.010$  is  little
probable.

\begin{figure} \epsfysize=370pt 
\epsfbox{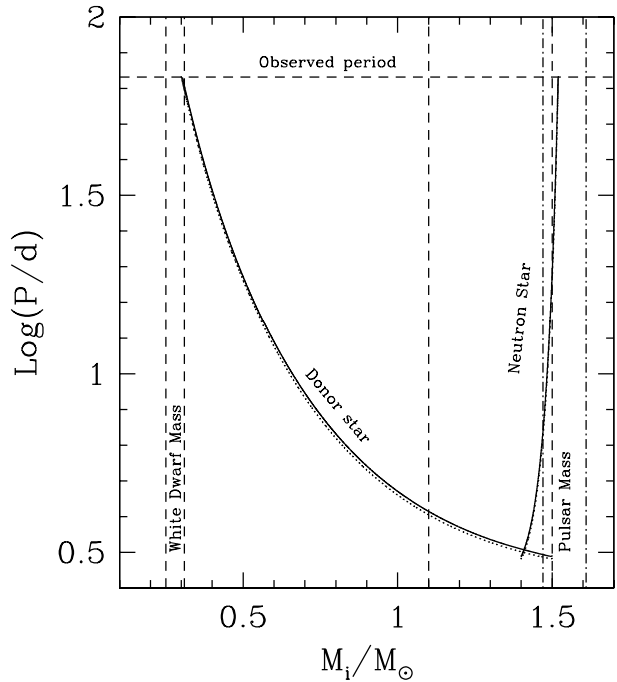}
\caption{Mass  vs.  Period for  the  evolutionary  sequences shown  in
Fig.~\ref{Fig:tracks}.  Dotted lines represent  the case of the system
with 1.50~\msol, a  metallicity of $Z=0.010$ and an  initial period of
$P_{i}=3.05$~days. Meanwhile,  solid lines stand  for the case  of the
same initial mass but a metallicity of $Z=0.020$ and an initial period
of  $P_{i}=3.10$~days.  Vertical  short   dashed  lines  on  the  left
represent the observed mass of the WD companion with its corresponding
error, while the  vertical short dashed lines on  the right correspond
to the mass of the NS also with its error bar. The other estimation of
the NS  mass derived employing  the orbital period-core  mass relation
(see  Section~\ref{sec:intro}) is shown  with vertical  dot-short dash
lines.   The horizontal long dashed line  represents the observed
orbital period (known with a very large degree of accuracy).   The
conditions considered  for an evolutionary sequence  to be acceptable
to account  for the characteristics  of the PSR~J1713+0747  system are
that both final masses (of the  WD and NS respectively) must be inside
the   error   bars   and   reach   the   observed   orbital   period.}
\label{Fig:masa_per} 
\end{figure}

The evolutionary tracks followed by  the normal star of the system are
shown in  Fig.~\ref{Fig:tracks} for each  metallicity value considered
here.  In both  cases the  star completes  its core  hydrogen burning,
evolves  to  the  red giant  branch  and  develops  an OCZ.  In  these
conditions the star overflows  its Roche lobe (RLOF) which corresponds
to   the   onset   of   the   initial  mass   transfer   episode.   In
Fig.~\ref{Fig:masa_per}  we show the  evolution of  the period  of the
system as a function of the  masses of the components of the pair. For
both  cases  the mass  of  the donor  star  finally  falls inside  the
1$\sigma$  error bar.  For  the case  of  the NS,  its  final mass  is
slightly larger than the one found by Splaver et al. (2005) but inside
the  error bar  of the  allowed mass  interval if  a  theoretical mass
radius  relation   is  considered.  In  view  of   its  present  large
uncertainty,  we  shall  consider  the   mass  value  for  the  NS  as
acceptable.

It  is important  to remark  that, in  both cases  we have  found that
during the  initial RLOF  the mass transfer  is stable and  the system
does not suffer  from a common envelope episode  (see Podsiadlowski et
al. 2002 and Benvenuto \& De Vito 2005 for similar results). After the
RLOF the star  contracts and diffusion carries hydrogen  inwards up to
the point at  which the bottom of the hydrogen  envelope is hot enough
to ignite  hydrogen in semi-degenerate  conditions (Althaus, Serenelli
\& Benvenuto 2001). Then a  thermonuclear flash occurs. In the case of
$Z=0.010$  the star  undergoes  two flashes  while  for $Z=0.020$  the
object experiences  only one. After flashes  the star cools  down as a
He~WD with  an outermost layer  rich in hydrogen. For  further details
see, e.g., Benvenuto \& De Vito (2004; 2005) (see also Sarna, Ergma \&
Gerskevits-Antipova   2000  and   Nelson  et   al.  2004).   The  main
characteristics  of the  WD final  cooling  tracks are  given in  
Tables~\ref{table:z0010} and  \ref{table:z0020} for the  cases of $Z=
0.010$ and $Z= 0.020$ respectively

\begin{figure} \epsfysize=320pt 
\epsfbox{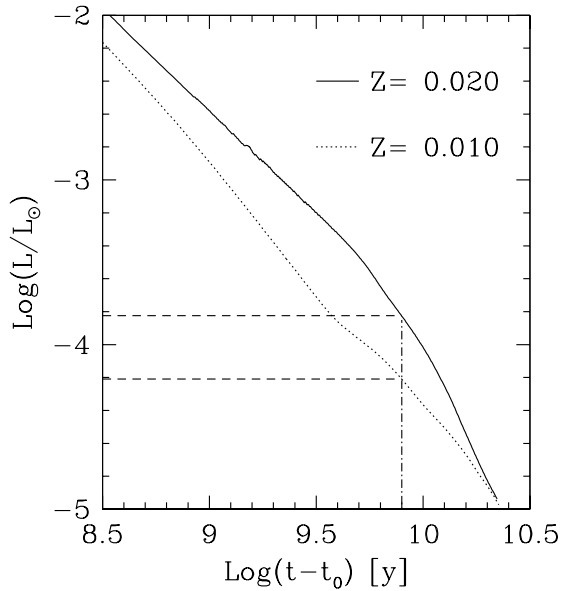} 
\caption{Evolution of  luminosity as a function of  time counted since
the end of  the initial mass transfer episode.  Solid and dotted lines
have the same meaning as in Fig.~\ref{Fig:masa_per}.  The vertical
dot-short dashed line represents  the characteristic timescale of the
MSP which,  because of evolutionary reasons, should  correspond to the
present conditions  of the He~WD. Horizontal short  dashed lines stand
for the luminosity the WD should have at the characteristic age of the
pulsar.} 
\label{Fig:cooling} 
\end{figure}

\begin{figure} 
\epsfysize=320pt 
\epsfbox{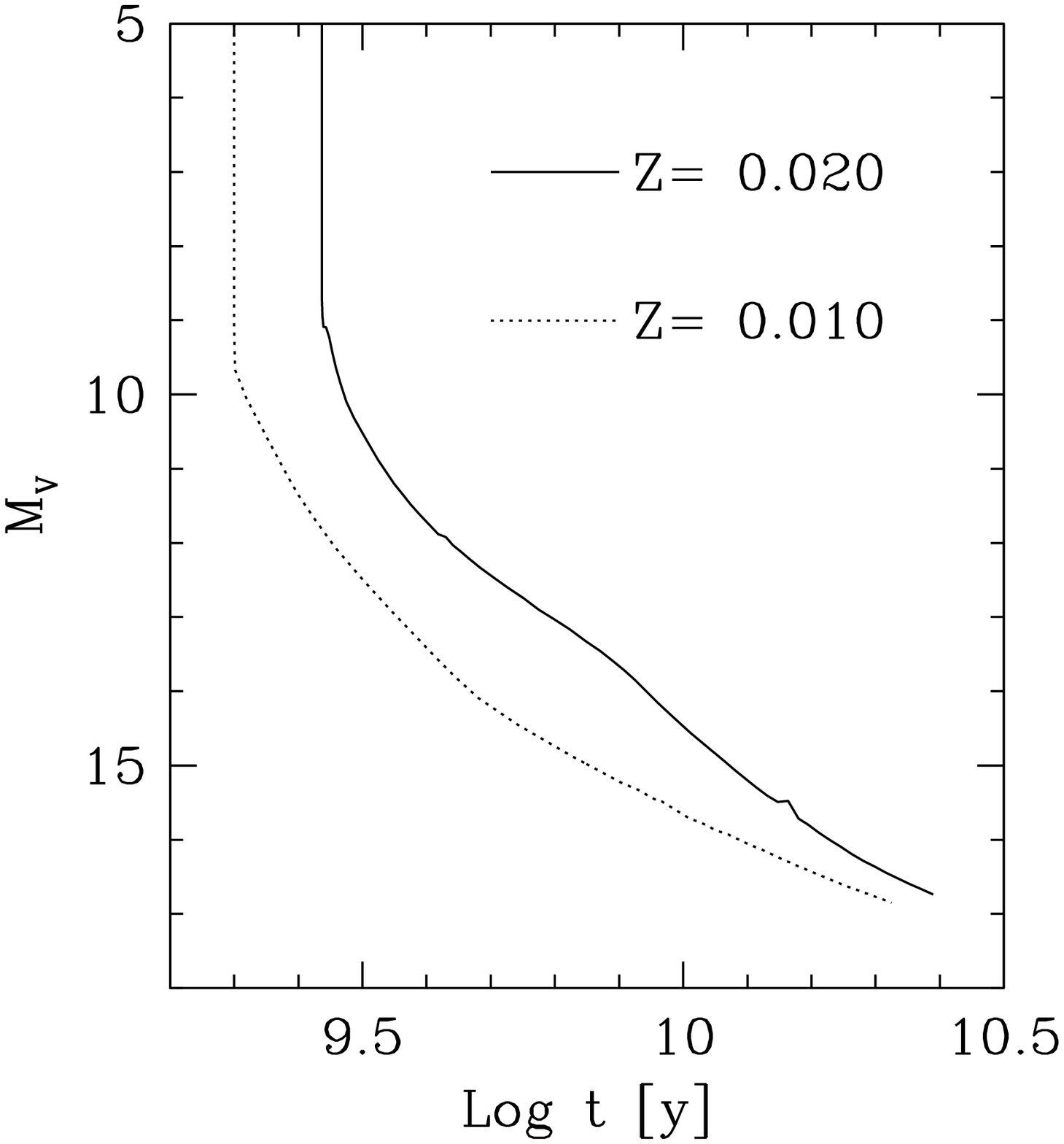} 
\caption{The evolution of the  absolute visual magnitude as a function
of the age of the star counted from the zero age main sequence on.}
\label{Fig:mv_t} 
\end{figure}

In Fig.~\ref{Fig:cooling}  we show the  luminosity of the models  as a
function of the age counted since  the end of the initial RLOF. Notice
that,  in spite  of  the  fact that  the  masses of  the  WDs of  both
considered cases  are very similar,  there is an  important difference
regarding the cooling  of the models. As models  with $Z=0.010$ evolve
faster  than  those  with  $Z=0.020$,   the  WD  is  predicted  to  be
fainter. Specifically, at  the characteristic age of the  MSP, for the
case of the evolutionary track with $Z=0.010$ ($Z=0.020$) the WD has a
luminosity     of    $\log{L/L_{\odot}}=     -4.210$~$(-3.825)$.    In
Fig.~\ref{Fig:mv_t}  we  show the  evolution  of  the absolute  visual
magnitude  as a  function of  the age  of the  star counted  since the
beginning of core hydrogen burning.

\begin{figure} 
\epsfysize=380pt 
\epsfbox{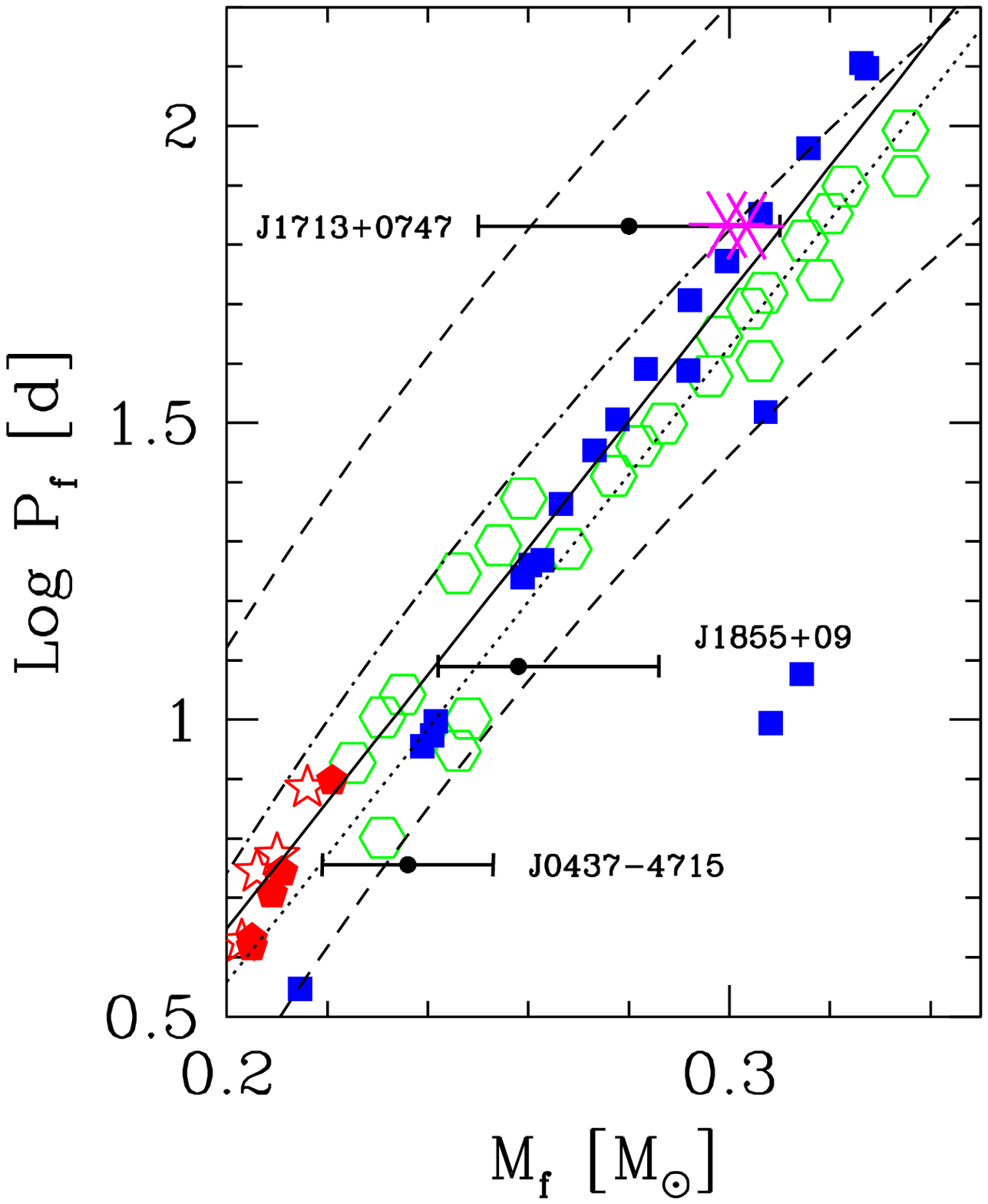} 
\caption{The mass vs.   orbital period relation for He~WDs  and the WD
in  PSR~J1713+0747  system.  Starred  hexagons  denote  the two  cases
considered here that account for  the characteristics of the system we
are studying here.  Solid (dotted) line denotes the relation predicted
by  Nelson  et  al.  (2004)  for  a  metallicity  value  of  $Z=0.020$
($Z=0.010$).   Dot-short  dash  line  depicts the  relation  given  in
Rappaport et al. (1995) together  with its uncertainty, shown in short
dash lines.  Filled squares, green  hollow hexagons, hollow  stars and
filled pentagons stand for data  from Benvenuto \& De Vito (2005) (for
$Z=0.020$); Podsiadlowski  et al. (2002) (for $Z=0.020$)  and Sarna et
al. (2000) for $Z= 0.020$  and $Z=0.010$ respectively. We also include
the  other  two  accurately  measured masses  (in  PSR~J0437-4715  and
PSR~B1855+09 systems) of low mass  WDs in binary systems together with
MSPs.}
\label{Fig:mfpf} 
\end{figure}

In Fig.~\ref{Fig:mfpf}  we show the final mass-orbital  period for low
mass He~WDs  in orbit  with MSPs including  the one  in PSR~J1713+0747
system  and the other  well determined  cases (in  PSR~J0437-4715 and
PSR~B1855+09 systems) together with  some recent calculations by Sarna
et al. (2000), Podsiadlowski et  al. (2002), Nelson et al. (2004), and
Benvenuto \& De Vito (2005)  and also the relations given by Rappaport
et al. (1995) and Nelson et al.(2004). The WD in PSR~J1713+0747 system
is the  most massive among the  three accurately measured  WDs of this
type.  While the  mean  value of  $M_{WD}=  0.28$~\msun is  marginally
compatible  with the  available calculations,  the 1$\sigma$  value of
$M_{WD}= 0.31$~\msun  is in nice agreement  with previous computations
as well as with those performed in this paper.

\begin{figure} 
\epsfysize=340pt 
\epsfbox{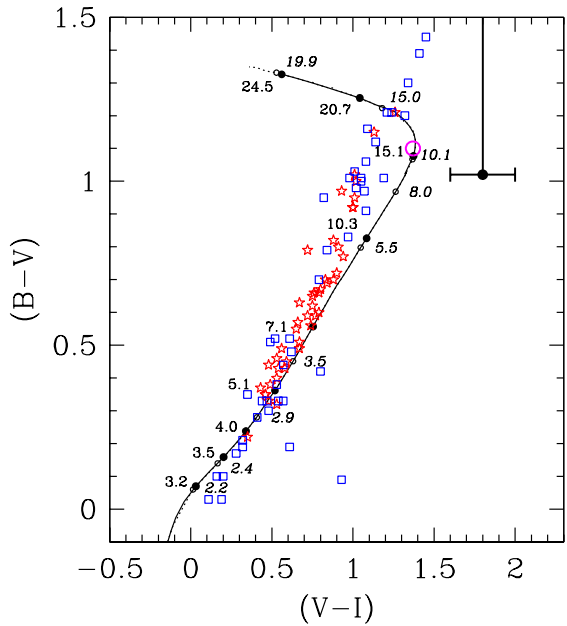} 
\caption{The  evolutionary tracks  of  the models  considered in  this
paper  as representative  of the  WD in  PSR~J1713+0747 system  in the
(B-V) vs.  (V-I) plane.  The  dotted (solid) line  stands for the
results with  $Z=0.010$ ($Z=0.020$).  Actually both  models yield
very  similar  colour-colour  curves.  Solid  circles  represent  the
conditions for the case  of $Z=0.020$ given in Table~\ref{table:z0020}
for which labels  to the left are their ages  (in Gyr). Hollow circles
stand for the conditions corresponding  to the case of $Z=0.010$ given
in Table~\ref{table:z0010} for which  labels to the right (in italics)
are  their ages  (in Gyr).  The best  fit corresponding  to the  WD in
PSR~J1713+0747  system  is shown  with  an  open  circle   located
approximately at (B-V)$=1.1$ and (V-I)$=1.4$, while the observational
data given  in Lundgren et al. (1996)  is shown with a  solid dot with
its  corresponding error bars.  For comparison  we have  also included
data corresponding to DA and non-DA WDs (denoted with hollow stars and
hollow squares) given by Bergeron,  Leggett \& Ruiz (2001) and Mc Cook
\& Sion (1987).   The value of (B-V) corresponding to  the best fit is
compatible  with the  lower limit  given  by Lundgren  et al.  (1996),
however (V-I) is too high for hydrogen atmospheres.}
\label{Fig:bv_vi} \end{figure}

\begin{figure} 
\epsfysize=350pt 
\epsfbox{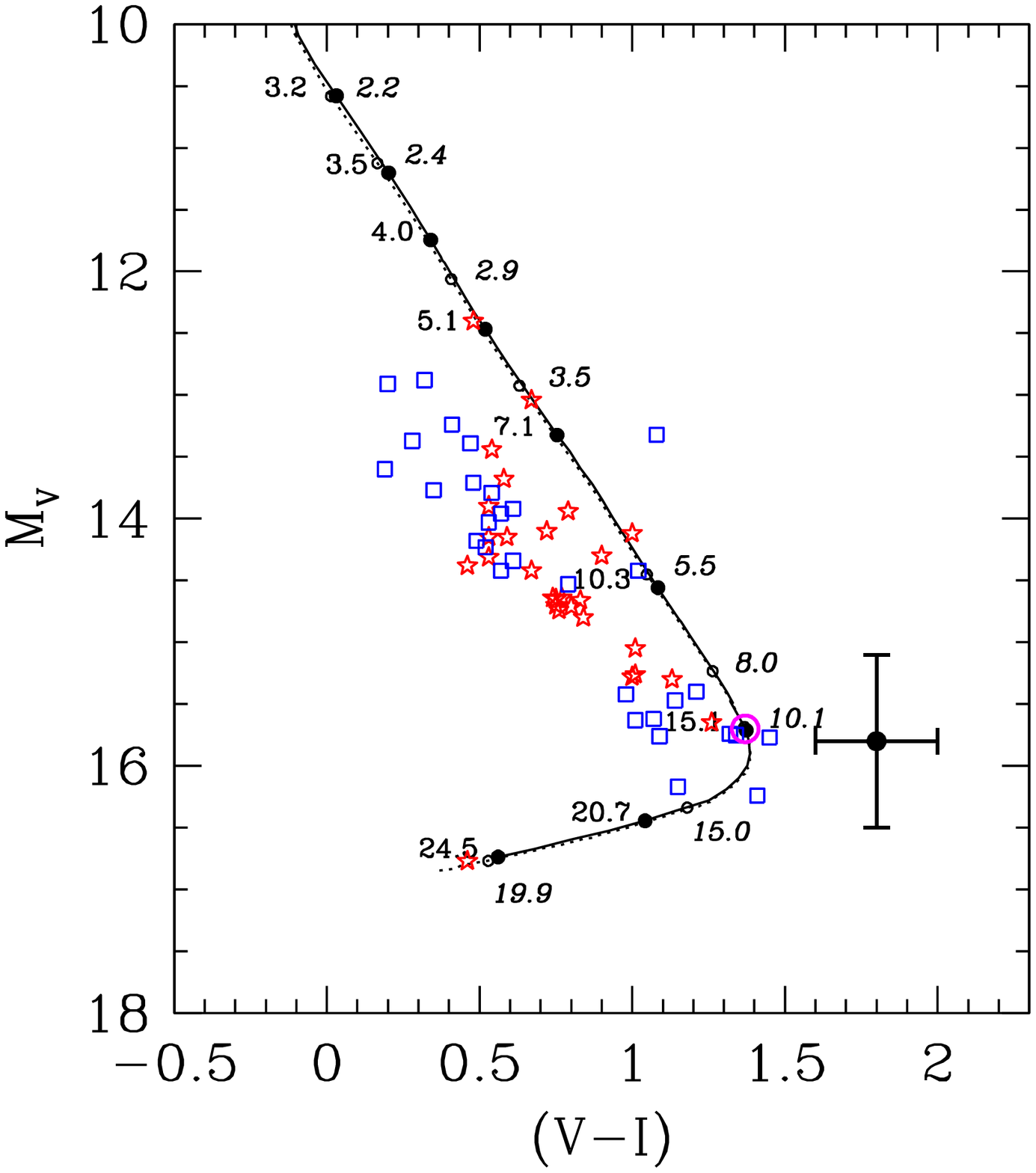} 
\caption{The  evolutionary tracks  of  the models  considered in  this
paper  as representative  of the  WD in  PSR~J1713+0747 system  in the
(V-I) vs. $M_V$ plane. Lines, labels and symbols have the same meaning
as in Fig.~\ref{Fig:bv_vi}.  Notice that, while the $M_V$ is very well
accounted  for by  our models,  (V-I) is  far higher  than  the values
predicted by almost pure hydrogen atmospheres.}
\label{Fig:mv_vi} 
\end{figure}

Having determined the possible  evolutionary sequences followed by the
normal,  donor star  in  PSR~J1713+0747  binary system,  we  are in  a
position  to   analyze  the  observational  appearance   of  the  last
contraction  toward  the   WD  regime.  In  Figs.~\ref{Fig:bv_vi}  and
\ref{Fig:mv_vi} we  show the computed  ($(B-V)$,$(V-I)$) colour-colour
and ($M_V$,$(V-I)$)  magnitude-colour diagrams.  Observations  of cool
WDs (symbols) have been plotted  for comparison. The cooling curves at
$Z=0.010$   (dotted   lines   on  Figs.   \ref{Fig:bv_vi}   and
\ref{Fig:mv_vi}) and $Z=0.020$ (solid lines in the same Figs.)
are  essentially  identical because  both  tracks  correspond to  very
similar  stellar  masses   ($M=0.302M_\odot$  within  1\%)  and
atmospheric   composition.  The  main   difference  between   the  two
evolutionary  solutions arises from  the cooling  ages (labels  on the
plots).

Since  gravitational settling  dominates  the chemical  stratification
after   a  relatively   short  cooling   time  (typically   0.1  Myr),
evolutionary  calculations predict  that  the WD  has mostly  hydrogen
floating  at  its  visible  surface (photosphere).  The  formation  of
molecular hydrogen becomes important as the star cools down. The H$_2$
CIA opacity is very strong  in the infrared and largely determines the
shape of  the spectral  energy distribution for  cool WD models.  As a
consequence of the molecular opacity, the $(V-I)$ colour becomes bluer
when the effective temperature falls  below 3600 K. The turnoff of the
WD  cooling sequences  is located  at $(V-I)  \approx 1.4$,  with $M_V
\approx 15.8$ and $(B-V) \approx 1.1$.

As  stated  in  the  introduction  (Section~\ref{sec:intro}),  optical
observations of the pulsar companion have been carried out by Lundgren
et  al. (1996)  (see Table~\ref{table:observa}).  The  timing parallax
distance    of    $1.1\pm0.1$~kpc    yields   a    distance    modulus
$m-M=10.2\pm0.2$. If we adopt  an interstellar absorption of $A_V=0.1$
as  a conservative estimate  (Burstein \&  Heiles 1982),  the absolute
visual magnitude  of the pulsar companion  is $M_V=15.7\pm0.4$.  Thus,
the observed companion is located near the turnoff of the evolutionary
sequences (where  $T_{eff}\approx 3600$~K) although  appears redder in
the     $(V-I)$      colour     (see     Figs.~\ref{Fig:bv_vi}     and
\ref{Fig:mv_vi}). Unfortunately, multicolour optical observations only
provide  a  lower limit  for  $(B-V)$ colour  such  that  it offers  a
relatively  poor constraint.  The observed  $(V-I)$ colour  is $\approx
0.4$~mag redder than  the cooling sequences. It is  worth noting that,
due  to the  CIA opacity,  $(V-I) \approx  1.5$ represents  an extreme
upper limit  for WDs with hydrogen-rich atmospheres  (e.g. Rohrmann et
al. 2002). It is therefore  somewhat disturbing that the $(V-I)$ index
observed  by  Lundgren  et  al.   (1996)  is  too  red  compared  with
rich-hydrogen envelope predictions. The  origin of this discrepancy is
unknown.  It could  be  due to  a  missing opacity  source  in the  WD
atmosphere  itself  or a  non  hydrogen-rich composition.   Additional
observations are necessary to disentangle this issue.

\begin{figure} 
\epsfysize=270pt 
\epsfbox{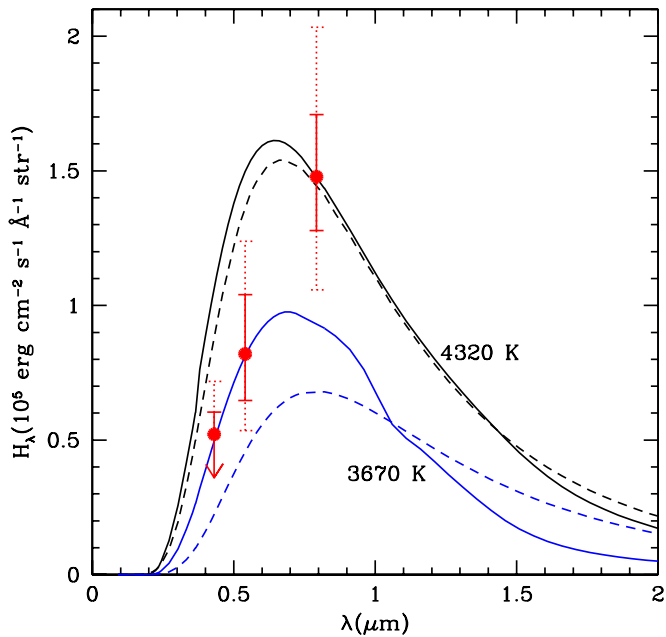} 
\caption{Energy   distribution   of    WD   model   atmospheres   with
$T_{eff}=4320$ and 3670 K (continuous lines) and the blackbody spectra
at these temperatures (dashed lines). The photometric observations are
represented  by   filled  circles.   Continuous   error  bars  include
uncertainties  in stellar  radius  and photometric  data only.  Dotted
error bars includes the parallax error also. }
\label{Fig:fluxt} \end{figure}

We   can   convert  the   observed   magnitudes   ($m$)  into   fluxes
($f_\lambda^m$)  and compare the  resulting energy  distributions with
those predicted  from our  model atmosphere calculations.  The average
flux  and the  magnitude for  each passband  is related  in  the form
$$
m = -2.5 \log f_\lambda^m - c_m \,,
$$ with $c_m$ being a  calibration constant. The Eddington flux at the
stellar surface $H_\lambda^m$ is evaluated from the relation
$$
f_\lambda^m = 4\pi (R/D)^2 H_\lambda^m
$$ where $R/D$ is the ratio of  the radius of the star to its distance
from Earth.   We use  the filter transmission  curves given  in Bessel
(1990),  the parallax  distance  of $D=1.1\pm  0.1$  kpc (from  pulsar
timing), averaged  values of the interstellar  extinction (Burstein \&
Heiles  1982), and  the radius  derived from  our  evolutionary models
$R=(1.90 \pm 0.05)\times 10^{-2}R_\odot$ (over a range of temperatures
$T_{eff}<5000$  K).  \footnote{The WD  radius  is  nearly constant  at
advanced   cooling  stages.}   The  computed   fluxes  are   shown  in
Fig. \ref{Fig:fluxt}.  These values confirm that  the binary companion
apparently exhibits  a strong  infrared flux {\em  excess} (or  a blue
deficiency).

The photometric  spectrum was fitted using  the evolutionary sequences
at  $Z=0.010$ and $Z=0.020$  (their surface  gravities differ  by less
than 0.1\%  for the  temperature range of  interest). The  results are
presented in Fig.~\ref{Fig:fluxt} and listed in Table~\ref{Table:atm}.
A good fit  to the energy distribution of the WD  companion can not be
achieved with a unique effective temperature. The flux at the $V$ band
can be  represented by  a model with  $T_{eff}=3670 \pm  280$~K, which
also fits the upper limit detected for the $B$ band.  However, for the
$I$ filter we obtain  $T_{eff}=4320\pm 180$~K.  These solutions, which
differ  quantitatively by  650~K, fall  within the  range  of previous
temperature  evaluations:  $T_{eff}=3700\pm 100$  K  (Lundgren et  al.
1996, based on a  temperature calibration of $(V-I)$ colour determined
by Monet  et al.  1992), $T_{eff}= 3430\pm  270$ K (Hansen  \& Phinney
1998, based on blackbody fits to HST data) and $T_{eff}=4250\pm 250$~K
(Sch\"onberner,  Driebe  \&   Bl\"ocker  2000,  from  evolutionary  WD
models).

\begin{figure*} 
\epsfysize=320pt 
\epsfbox{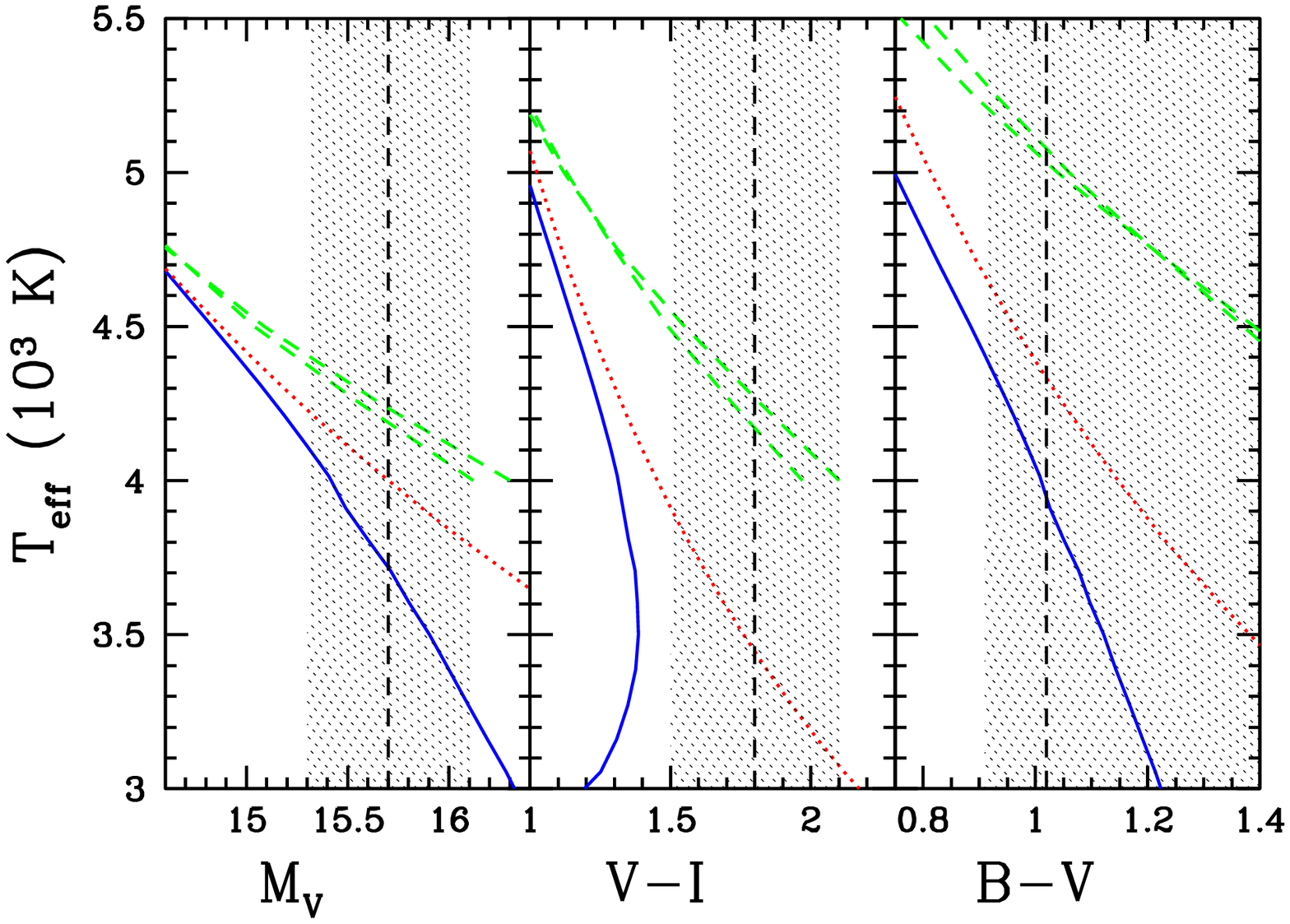}
\caption{Effective    temperature   vs.    absolute    magnitude   and
colours. Solid  lines denotes the cooling WD  track with hydrogen-rich
atmospheres predicted  for the millisecond pulsar  companion.  The
solid $V-I$  curve turns to  the left at  low temperatures due  to the
molecular  hydrogen   opacity.   Dotted  lines   represent  blackbody
solutions, and  dashed lines  correspond to helium-pure  atmosphere at
$\log  g=7.0$  and  7.5   taken  from  Bergeron,  Saumon  \&  Wesemael
(1995). Vertical  dashed lines correspond  to the observed  values for
the companion of pulsar J1713+0747 (the WD absolute magnitude is based
on the distance inferred from  the dispersion measure). We denote with
a  shade  area  the  range  compatible  with  the  photometric  data.}
\label{Fig:t_mbvi}
\end{figure*}

The relation between photometric data and effective temperature of the
WD  can  be appreciated  in  Fig.~\ref{Fig:t_mbvi}. There,  broad-band
colours  and absolute visual  magnitude of  the observed  star (dashed
vertical   lines)   are  compared   with   predictions  from   present
evolutionary   calculations  (solid  line),   blackbody  approximation
(dotted line), and pure  helium broad-band colours of Bergeron, Saumon
\&  Wesemael  (1995)  for $\log  g  =  7.0$  and 7.5  (dashed  lines).
Absolute magnitudes  derived for helium models were  evaluated using a
radius  of  $R=1.9\times  10^{-2}R_\odot$ and  bolometric  corrections
provided  by those  authors. The  $(V-I)$ broad-band  colour  yields a
temperature estimate  of 3450~K when a blackbody  spectrum is assumed,
which  is considerably  lower than  the temperature  derived  from the
$M_V$ fit ($\approx 4000$~K). Hydrogen-rich atmospheres of WDs deviate
strongly  from the  blackbody approximation  at low  $T_{eff}$  as the
opacities become dominated by molecular hydrogen. The observed $(V-I)$
colour   is   far   off    the   hydrogen   sequence   as   shown   in
Fig.~\ref{Fig:t_mbvi} (middle panel),  whereas the $M_V$ fit indicates
a temperature of approximately 3700~K. On the other hand, the observed
$BVI$ magnitudes  can be understood  with a helium-pure  atmosphere if
$T_{eff}\approx  4250\pm  300$~K.  All  cases fall  within  the  limit
observed for the $(B-V)$ colour.

This exercise shows that the effective temperature of the WD companion
of  PSR J1713+0747  should be  between 3700  and 4300~K.   Taking into
account  the uncertainties in  the photometric  data fits,  we finally
adopt $T_{eff}= 4000\pm 400$~K as a conservative solution based on HST
observations. Consequently, using the $T_{eff}-$radius relation of our
evolutionary  calculations, we  determine  a luminosity  in the  range
$-3.92    >    \log   (L/L_\odot)    >    -4.29$.     As   shown    in
Fig.~\ref{Fig:tracks},  the detected optical  counterpart lies  on the
calculated cooling  tracks.  Furthermore, for  the case of  $Z= 0.010$
the  cooling age  deduced from  our  models is  $9 \pm  2$~Gyr. If  we
consider that the star ends its  initial RLOF with an age of 1.99 Gyr,
we  find  that the  WD  has  spent $7  \pm  2$~Gyr  in evolving  since
detachment,  which  is  in   excellent  agreement  with  the  pulsar's
characteristic age!  However, for $Z= 0.020$ our  models indicate that
the  star  needs  $13.5 \pm  2$~Gyr  in  evolving  up to  the  present
stage. For this metallicity,  detachment occurs at 2.73~Gyr. Then, the
star would have needed $10.8\pm2$~Gyr which is a larger timescale than
the characteristic  age of the  pulsar.  This fact  indicates that
the donor star very likely has a metallicity $Z\approx 0.010$.

\begin{centering}
\begin{table*}
\caption{\label{table:z0010}  Selected  evolutionary  stages  for  the
system  with   a  metallicity  of   $Z=  0.010$,  initial   period  of
3.05~days. We  tabulate the  age, effective temperature,  logarithm of
the luminosity, logarithm of the central temperature, logarithm of the
central density,  the mass fraction  embraced in the  outer convective
zone,  the  fractional amount  of  hydrogen  in  the whole  star,  the
gravitational  acceleration at  the stellar  surface,  absolute visual
magnitude,  colour indices and  the bolometric  correction. Quantities
without specified units  are given in cgs. The  final mass and orbital
period values for  this system are $M_f= 0.2994507  M_\odot$ and $P_f=
68.288$~days respectively.}
\begin{tabular}{rrrrrrrrrr}
\hline
\hline
$Age/Gyr$ & $T_{eff}$ & $\log{(L/L_\odot)}$ & $\log{T_C}$  & $\log{\rho_C}$  & $10^3 q_{OCZ}$ & $10^4 M_H/M_*$ & $\log{g}$ & $M_V$   & \\
          &  $(U-B)$  &  $(B-V)$            &  $(V-R)$     &  $(V-K)$        &  $(R-I)$       &  $(J-H)$       &  $(H-K)$  & $(K-L)$ & $BC$ \\
\hline
 2.245 &  10772 & -2.167 & 7.263 &  5.647 &  0.000 &  7.492 &  7.165 & 10.582 & \\
       & -0.404 &  0.060 & 0.010 & -0.143 &  0.004 &  0.047 & -0.116 & -0.036 & -0.413 \\
 2.425 &   9424 & -2.439 & 7.160 &  5.662 &  0.000 &  7.473 &  7.205 & 11.124 & \\
       & -0.452 &  0.140 & 0.083 &  0.153 &  0.083 &  0.090 & -0.097 & -0.033 & -0.274 \\
 2.872 &   7622 & -2.858 & 6.999 &  5.678 &  0.001 &  7.450 &  7.256 & 12.062 & \\
       & -0.517 &  0.280 & 0.201 &  0.698 &  0.206 &  0.178 & -0.059 & -0.012 & -0.166 \\
 3.515 &   6301 & -3.223 & 6.866 &  5.687 &  0.016 &  7.430 &  7.290 & 12.924 & \\
       & -0.390 &  0.451 & 0.312 &  1.221 &  0.319 &  0.267 & -0.011 &  0.035 & -0.114 \\
 5.515 &   4812 & -3.756 & 6.659 &  5.697 &  9.747 &  7.406 &  7.355 & 14.450 & \\
       &  0.060 &  0.798 & 0.523 &  2.071 &  0.525 &  0.366 &  0.057 &  0.170 & -0.308 \\
 8.001 &   4184 & -4.021 & 6.480 &  5.702 &  9.747 &  7.405 &  7.377 & 15.234 & \\
       &  0.233 &  0.968 & 0.631 &  2.085 &  0.632 &  0.138 & -0.021 &  0.387 & -0.430 \\
10.118 &   3750 & -4.220 & 6.349 &  5.705 &  7.303 &  7.407 &  7.385 & 15.684 & \\
       &  0.302 &  1.067 & 0.690 &  1.558 &  0.677 & -0.176 & -0.167 &  0.763 & -0.382 \\
14.967 &   2995 & -4.620 & 6.109 &  5.709 &  3.907 &  7.406 &  7.395 & 16.337 & \\
       &  0.489 &  1.223 & 0.700 &  0.181 &  0.480 & -0.386 & -0.438 &  1.738 & -0.035 \\
19.919 &   2502 & -4.936 & 5.935 &  5.710 &  1.853 &  7.405 &  7.399 & 16.768 & \\
       &  0.654 &  1.331 & 0.594 & -0.876 & -0.065 & -0.269 & -0.734 &  2.557 &  0.324 \\
\hline
\end{tabular}     	    	    	    	        	 	    	   	   
\end{table*}	   	    	    	    	        	 	    	   	   
\end{centering}

\begin{centering}
\begin{table*}
\caption{\label{table:z0020}  Selected  evolutionary  stages  for  the
system  with   a  metallicity  of   $Z=  0.020$,  initial   period  of
3.10~days.     Columns    have    the     same    meaning     as    in
Table~\ref{table:z0010}. The final mass  and orbital period values for
this  system  are  $M_f=  0.3033765 M_\odot$  and  $P_f=  67.762$~days
respectively.}
\begin{tabular}{rrrrrrrrrr}
\hline
\hline
$Age/Gyr$ & $T_{eff}$ & $\log{(L/L_\odot)}$ & $\log{T_C}$  & $\log{\rho_C}$  & $10^3 q_{OCZ}$ & $10^4 M_H/M_*$ & $\log{g}$ & $M_V$   & \\
          &  $(U-B)$  &  $(B-V)$            &  $(V-R)$     &  $(V-K)$        &  $(R-I)$       &  $(J-H)$       &  $(H-K)$  & $(K-L)$ & $BC$ \\
\hline
 3.191 &  10582 & -2.176 & 7.194 &  5.673 &  0.000 & 16.696 &  7.150 & 10.579 & \\
       & -0.404 &  0.070 & 0.018 & -0.110 &  0.013 &  0.052 & -0.114 & -0.036 & -0.387\\
 3.542 &   9141 & -2.480 & 7.066 &  5.686 &  0.000 & 16.428 &  7.199 & 11.203 & \\
       & -0.466 &  0.159 & 0.100 &  0.227 &  0.102 &  0.100 & -0.093 & -0.031 & -0.253\\
 4.010 &   8104 & -2.721 & 6.969 &  5.694 &  0.001 & 16.194 &  7.230 & 11.745 & \\
       & -0.514 &  0.238 & 0.168 &  0.537 &  0.172 &  0.150 & -0.071 & -0.021 & -0.192\\
 5.059 &   6895 & -3.035 & 6.863 &  5.701 &  0.006 & 15.868 &  7.264 & 12.467 & \\
       & -0.478 &  0.362 & 0.256 &  0.966 &  0.263 &  0.226 & -0.036 &  0.008 & -0.128\\
 7.054 &   5778 & -3.376 & 6.772 &  5.706 &  0.140 & 15.531 &  7.298 & 13.323 & \\
       & -0.260 &  0.557 & 0.374 &  1.483 &  0.380 &  0.301 &  0.020 &  0.069 & -0.132\\
10.272 &   4713 & -3.788 & 6.635 &  5.713 & 11.219 & 15.300 &  7.356 & 14.558 & \\
       &  0.090 &  0.826 & 0.541 &  2.116 &  0.543 &  0.364 &  0.044 &  0.194 & -0.336\\
15.114 &   3707 & -4.235 & 6.330 &  5.719 &  7.882 & 15.274 &  7.386 & 15.712 & \\
       &  0.320 &  1.077 & 0.695 &  1.480 &  0.678 & -0.202 & -0.192 &  0.818 & -0.372\\
20.711 &   2851 & -4.701 & 6.072 &  5.723 &  3.363 & 15.257 &  7.396 & 16.444 & \\
       &  0.535 &  1.254 & 0.679 & -0.106 &  0.363 & -0.368 & -0.507 &  1.959 &  0.061\\
24.505 &   2520 & -4.918 & 5.944 &  5.724 &  1.923 & 15.257 &  7.399 & 16.737 & \\
       &  0.647 &  1.326 & 0.599 & -0.828 & -0.038 & -0.273 & -0.713 &  2.517 &  0.309\\
\hline	 	   	    	    	    	    	        			
\end{tabular}     	    	    	    	        	 	    	   	   
\end{table*}	   	    	    	    	        	 	    	   	   
\end{centering}

\begin{table*}
\caption{\label{Table:atm}  WD  model   atmospheres  for  the  optical
observations by Lundgren et al. (1996).}
\begin{tabular}{ccccccc}
\hline
\hline
$T_{eff}$(K) & $\log{(L/L_\odot)}$ & $\log g$ & $M_V$ & $(B-V)$ & $(V-I)$ & BC \\
\hline
 4320  & -3.95 & 7.373 & 15.032 & 0.932 & 1.218  & -0.415 \\
 3670  & -4.23 & 7.387 & 15.689 & 1.085 & 1.377  & -0.364 \\ 
\hline
\end{tabular}
\end{table*}

%---------------------------------------------------------------------
\section{Discussion and Conclusions} \label{sec:discu}

In this paper we have computed  a set of binary evolutionary tracks in
order to  account for the  main characteristics of  the PSR~J1713+0747
system. We  did so  with a binary  stellar evolution code  including a
non-grey treatment of the atmospheric layers.

Thanks to the high accuracy  of the timing of this millisecond pulsar,
it has been possible to measure the general relativistic Shapiro delay
effect to  derive separate  values for the  masses of the  white dwarf
(WD)  and the  neutron  star (NS).  Starting  out from  this fact,  we
performed  a  numerical  experiment  in  which we  computed  the  full
evolution of the system in order to account for the masses and orbital
period of the system and {\it then} considered the evolutionary status
of the  WD.  Notice that  the pulsar should  have a characteristic
age equal to  the age of the  WD counted since the end  of the initial
Roche   lobe  overflow   (see  Section~\ref{sec:intro}   for  further
details). Thus, at the characteristic age of the pulsar, the WD should
resemble the object observed by Lundgren et al.  (1996).

We found that a system with initial masses of $M= 1.5$~\msol, $M_{NS}=
1.4$~\msun for the normal (donor) star and NSs (respectively) in which
the NS is able to accrete  a fraction $\beta= 0.10$ of the transferred
mass  (the mass  lost  from the  system  is assumed  to  carry away  a
specific angular  momentum equal to that  of the NS)  accounts for the
PSR~J1713+0747 system  if $Z= 0.010$  and $P_i= 3.05$~d or  $Z= 0.020$
and $P_i=  3.10$~d. We  did not   try to compute  detailed models
with lower initial  masses, because in such a case,  the WD would need
to spend a  time in excess of  the age of the Universe  in evolving to
the  observed configuration.  Also, at  longer initial  periods models
would undergo a common envelope episode because of the very deep outer
convective  zone of  the donor  star at  the onset  of the  Roche lobe
overflow.

In  our opinion,  it  is remarkable  that  the model  with $Z=  0.010$
accounts  for the masses,  the orbital  period of  the system  and the
brightness of the WD observed by Lundgren et al. (1996) at the correct
age. We consider it as a successful fit which, in turn, indicates that
the stellar  evolution processes relevant  for the present  system are
fairly well known.  Notice that we would have been able to predict
the  absolute magnitude  of the  WD  in PSR~J1713+0747  system with  a
reasonable accuracy.  Our models fit the main characteristics of
the WD at the age corresponding to $\tau_{PSR}=8$~Gyr given by Splaver
et al.  (2005).  Consequently,  we do not  find evidence for  a faster
cooling as in Hansen \& Phinney (1998).

However,  as our models  predict pure  hydrogen atmospheres,  they are
unable to  reproduce the  colours measured by  Lundgren et  al. (1996)
within the error bars of  the observations. In any case, we
should  remark that  this  discrepancy is  moderate  since theory  and
observations  differ in  less  than $3\sigma$.   Moreover,  the
available photometric data  is not yet accurate enough  to determine a
precise effective temperature  and atmospheric chemical composition of
the  WD companion  of the  pulsar.   In  order to  get a  deeper
understanding of the evolution of this system it would be essential to
perform new spectroscopic and photometric observations.  Particularly,
infra-red observations  can offer much extra information  and may help
to constrain this binary further.

%---------------------------------------------------------------------
\section{acknowledgments}

OGB has been supported by FONDAP Center for Astrophysics 1501003. RDR
acknowledges partial support from SeCyT-UNC 123/04 and CONICET 691/04. 

%---------------------------------------------------------------------

\bsp

\label{lastpage}


\begin{thebibliography}{99}

\bibitem[Althaus  et  al.(2001)]{2001MNRAS.323..471A} Althaus,  L.~G.,
Serenelli, A.~M., \& Benvenuto, O.~G.\ 2001, \mnras, 323, 471
 
\bibitem[Alpar et al.(1982)]{1982Natur.300..728A} Alpar, M.~A., Cheng,
A.~F., Ruderman, M.~A., Shaham, J.\ 1982, \nat, 300, 728

\bibitem[Benvenuto  \& De  Vito(2003)]{2003MNRAS.342...50B} Benvenuto,
O.~G., \& De Vito, M.~A.\ 2003, \mnras, 342, 50

\bibitem[Benvenuto   De   Vito(2004)]{2004MNRAS.352..249B}  Benvenuto,
O.~G., De Vito, M.~A.\ 2004, \mnras, 352, 249

\bibitem[Benvenuto De Vito(2005)]{}  Benvenuto, O.~G., De Vito, M.~A.\
2005, \mnras, 362, 891

\bibitem[]{} Bergeron,  P., Saumon, D., Wesemael, F.  1995, \apj, 443,
764

\bibitem[Bergeron  et  al.(2001)]{2001ApJS..133..413B}  Bergeron,  P.,
Leggett, S.~K., Ruiz, M.~T.\ 2001, \apjs, 133, 413

\bibitem[Bessell(1990)]{1990PASP..102.1181B}   Bessell,  M.~S.\  1990,
\pasp, 102, 1181

\bibitem[]{} Borysow A., Jorgensen U. G., Fu Y., 2001, JQSRT, 68, 235

\bibitem[]{} Burstein, D., Heiles, C., 1982, \aj, 87, 1165

\bibitem[Camilo et al.(1994)]{1994ApJ...437L..39C} Camilo, F., Foster,
R.~S., Wolszczan, A.\ 1994, \apjl, 437, L39

\bibitem[Foster  et   al.(1993)]{1993ApJ...410L..91F}  Foster,  R.~S.,
Wolszczan, A., Camilo, F.\ 1993, \apjl, 410, L91

\bibitem[]{}  Gaur  V. P.,  Tripathi B.  M., Joshi  G.  C., Pande
M. C., 1988, ApSS, 147, 107

\bibitem[]{} Gustafsson M., Frommhold L., 2001, \apj, 546, 1168

\bibitem[Hansen  Phinney(1998)]{1998MNRAS.294..569H} Hansen, B.~M.~S.,
\& Phinney, E.~S.\ 1998, \mnras, 294, 569

\bibitem[]{}  Harris G. J.,  Lynas-Gray A. E., Miller S., Tennyson
J., 2004, \apj, 617, L143

\bibitem[]{} Jorgensen  U. G., Hammer D., Borysow  A., Falkesgaard J.,
2000, \aap, 361, 283

\bibitem[Kaspi   et   al.(1994)]{1994ApJ...428..713K}  Kaspi,   V.~M.,
Taylor, J.~H., Ryba, M.~F.\ 1994, \apj, 428, 713

\bibitem[Lundgren et  al.(1996)]{1996ASPC..105..497L} Lundgren, S.~C.,
Foster, R.~S., Camilo, F.\  1996, ASP Conf.~Ser.~105: IAU Colloq.~160:
Pulsars: Problems and Progress, 105, 497

\bibitem[McCook Sion(1987)]{1987ApJS...65..603M}  McCook, G.~P., Sion,
E.~M.\ 1987, \apjs, 65, 603

\bibitem[]{} Monet D.  G., Dahn C. C., Vrba F. J.,  Harris H. C., Pier
J. R., Luginbuhl C. B., Ables H. D., 1992, AJ, 103, 638

\bibitem[Nelson  et   al.(2004)]{2004ApJ...616.1124N}  Nelson,  L.~A.,
 Dubeau, E., MacCannell, K.~A.\ 2004, \apj, 616, 1124

\bibitem[Podsiadlowski        et       al.(2002)]{2002ApJ...565.1107P}
Podsiadlowski, P., Rappaport, S., Pfahl, E.~D.\ 2002, \apj, 565, 1107

\bibitem[Rappaport  et al.(1983)]{1983ApJ...275..713R}  Rappaport, S.,
Verbunt, F., \& Joss, P.~C.\ 1983, \apj, 275, 713
 
\bibitem[Rappaport  et al.(1995)]{1995MNRAS.273..731R}  Rappaport, S.,
Podsiadlowski,  P., Joss,  P.~C., Di  Stefano, R.,  \& Han,  Z.\ 1995,
\mnras, 273, 731
 
\bibitem[Rohrmann(2001)]{2001MNRAS.323..699R}  Rohrmann,  R.~D.\ 2001,
\mnras, 323, 699

\bibitem[Rohrmann et  al.(2002)]{2002MNRAS.335..499R} Rohrmann, R.~D.,
Serenelli, A.~M., Althaus, L.~G., Benvenuto, O.~G.\ 2002, \mnras, 335,
499

\bibitem[]{} Sarna,  M.~J., Ergma, E.,  Gerskevits-Antipova, J.\ 2000,
\mnras, 316,84

\bibitem[]{} Sch\"onberner  D., Driebe  T., Bl\"ocker T.,  2000, A\&A,
356, 929

\bibitem[Serenelli   et   al.(2001)]{2001MNRAS.325..607S}   Serenelli,
A.~M.,  Althaus,  L.~G.,  Rohrmann,  R.~D.,  Benvenuto,  O.~G.\  2001,
\mnras, 325, 607

\bibitem[Splaver  et  al.(2005)]{2005ApJ...620..405S} Splaver,  E.~M.,
Nice, D.~J., Stairs, I.~H.,  Lommen, A.~N., Backer, D.~C.\ 2005, \apj,
620, 405

\bibitem[Stairs(2004)]{2004Sci...304..547S}   Stairs,   I.~H.\   2004,
Science, 304, 547

\bibitem[Stancil(1994)]{1994ApJ...430..360S}   Stancil,  P.~C.\  1994,
\apj, 430, 360

\bibitem[Thorsett \& Chakrabarty(1999)]{1999ApJ...512..288T} Thorsett,
S.~E., \& Chakrabarty, D.\ 1999, \apj, 512, 288

\bibitem[van  Straten et al.(2001)]{2001Natur.412..158V}  van Straten,
W.,  Bailes,  M.,  Britton,  M.,  Kulkarni,  S.~R.,  Anderson,  S.~B.,
Manchester, R.~N., Sarkissian, J.\ 2001, \nat, 412, 158

\end{thebibliography}
\end{document}